# Using Empirical Trajectory Data to Design Connected Autonomous Vehicle Controllers for Traffic Stabilization

**Yujie Li**
Graduate Research Assistant, Center for Connected and Automated Transportation (CCAT), and Lyles School of Civil Engineering, Purdue University, West Lafayette, IN, 47907.
Email: li2804@purdue.edu
ORCID #: 0000-0002-0656-4603

**Sikai Chen\***
Postdoctoral Research Fellow, Center for Connected and Automated Transportation (CCAT), and Lyles School of Civil Engineering, Purdue University, West Lafayette, IN, 47907.
Email: chen1670@purdue.edu; and
Visiting Research Fellow, Robotics Institute, School of Computer Science, Carnegie Mellon University, Pittsburgh, PA, 15213.
Email: sikaichen@cmu.edu
ORCID #: 0000-0002-5931-5619
(Corresponding author)

**Runjia Du**
Graduate Research Assistant, Center for Connected and Automated Transportation (CCAT), and Lyles School of Civil Engineering, Purdue University, West Lafayette, IN, 47907.
Email: du187@purdue.edu
ORCID #: 0000-0001-8403-4715

**Paul (Young Joun) Ha**
Graduate Research Assistant, Center for Connected and Automated Transportation (CCAT), and Lyles School of Civil Engineering, Purdue University, West Lafayette, IN, 47907.
Email: ha55@purdue.edu
ORCID #: 0000-0002-8511-8010

**Jiqian Dong**
Graduate Research Assistant, Center for Connected and Automated Transportation (CCAT), and Lyles School of Civil Engineering, Purdue University, West Lafayette, IN, 47907.
Email: dong282@purdue.edu
ORCID #: 0000-0002-2924-5728

**Samuel Labi**
Professor, Center for Connected and Automated Transportation (CCAT), and Lyles School of Civil Engineering, Purdue University, West Lafayette, IN, 47907.
Email: labi@purdue.edu
ORCID #: 0000-0001-9830-2071

## ABSTRACT

Emerging transportation technologies offer unprecedented opportunities to improve the efficiency of the current transportation system in the US from the perspectives of energy consumption, congestion, and emissions. One of these technologies is connected and autonomous vehicles (CAVs). With the prospective duality of operations of CAVs and human driven vehicles in the same roadway space (also referred to as a mixed stream), CAVs are expected to address a variety of traffic problems particularly those that are either caused or exacerbated by the heterogeneous nature of human driving. In efforts to realize such specific benefits of CAVs in mixed-stream traffic, it is essential to understand and simulate the behavior of human drivers in such environments, and microscopic traffic flow (MTF) models can be used to carry out this task. By helping to comprehend the fundamental dynamics of traffic flow, MTF models serve as a powerful approach to assess the impacts of such flow in terms of safety, stability, and efficiency. In this paper, we seek to calibrate MTF models based on empirical trajectory data as basis of not only understanding traffic dynamics such as traffic instabilities, but ultimately using CAVs to mitigate stop-and-go wave propagation. The paper therefore duly considers the heterogeneity and uncertainty associated with human driving behavior in order to calibrate the dynamics of each HDV. Also, the paper designs the CAV controllers based on the microscopic HDV models that are calibrated in real time. The data for the calibration is from the Next Generation SIMulation (NGSIM) trajectory datasets. The results are encouraging, as they indicate the efficacy of the designed controller to significantly improve not only the stability of the mixed traffic stream but also the safety of both CAVs and HDVs in the traffic stream. The paper's results can therefore help relieve phantom traffic jams that are caused by irrational or spontaneous driving patterns of human drivers, which has been identified in the literature as one of the main causes of traffic congestion. Overall, the paper's results are essential for effective real-world deployment of CAV controllers in mixed traffic environments during the CAV transition era.

**Keywords:** Mixed traffic streams, Model calibration, Phantom traffic jams, Connected and autonomous vehicles.

## INTRODUCTION

New technologies in transportation, fueled largely by information and communication science, materials science, and artificial intelligence, include vehicle autonomy, vehicle connectivity, electric propulsion, shared mobility, and airborne personal transport. The emergence of these technologies continue to offer unparalleled opportunities to improve travel efficiency, enhance safety, and reduce energy consumption, congestion, and emissions (Federal Highway Administration, 2018; U.S. Department of Transportation, 2018; Schrank et al., 2019; Chen, 2019; Ha et al., 2020a; Du et al., 2020). There are extensive studies that address the environmental impacts of new or merging transportation systems in the context of sustainable mobility (Chen et al., 2020; Dong et al., 2020; Li et al., 2019; Madireddy et al., 2011; Zheng et al., 2011). Connected and autonomous vehicles (CAVs) in particular, have received tremendous interest recently. In the CAV transition era where both CAVs and human driven vehicles (HDVs) will share the same roadway space (this situation is often referred to as a mixed stream), the debilitating impacts of human driving to traffic efficiency, currently taken for granted and inevitable, will become all too obvious. There exists a variety of traffic problems that are either caused or amplified by the heterogeneous nature of human driving. The term "heterogeneous" is used rather charitably, for human driving is often erratic and irrational, as humans behind the wheel have been known to engage in non-driving activities (texting, eating, applying facial make-up, chatting) that causes irregular, irrational or spontaneous driving patterns, thereby slowing the traffic flow or rendering it anomalous, and thereby making it more susceptible to crashes.

The traffic instability and propagation of stop-and-go waves caused by such irregular behavior of human driving can trigger severe congestion even without physical bottlenecks (Gunter et al., 2019; Wu et al., 2018). This non-bottleneck related congestion is referred as *phantom traffic jams* (Helbing, 2001; Orosz et al., 2009). A number of experimental studies have reproduced the degradation in traffic string stability that leads to such phantom jams (Jiang et al., 2018; Stern et al., 2018; Sugiyamal et al., 2008) but did not provide explicit solutions to avoid these jams. Fortunately, with the advent of connected and autonomous vehicles (CAVs), there seems to be a promising solution to this traffic problem. With vehicle automation and connectivity-aided communication, the vehicle is afforded enhanced awareness of its surrounding conditions (Ha et al., 2020b; Dong et al., 2020a). It has been posited that CAVs can help reduce congestion, increase safety, improve productivity, and increase the capacity of existing transportation facilities (Talebpour and Mahmassani, 2016; Dong et al., 2020). This is consistent with the anticipation that CAVs can help resolve some longstanding transportation engineering problems as stated in recent publications by USDOE (Administration, 2017). Specifically, in effort to mitigate phantom traffic jams, researchers have determined that the flow of an entire traffic string can be stabilized by controlling the behavior of certain vehicles in the platoon (Stern et al., 2018; Wu et al., 2018). CAVs introduced into the existing traffic stream can enable precise control considering the overall traffic conditions and models of the human drivers, thereby dampening the shockwave. In previous work (Li et al., 2020), it has been found that in a mixed traffic stream, the HDV driver exhibits behavioral heterogeneity and perception-reaction time delay that impairs the capability of AVs to stabilize traffic flows. Hence, to implement the framework proposed, a real-time calibration of human-driven vehicles (HDVs) model capturing heterogeneity becomes essential to CAV implementation in the real world.

In efforts to realize and quantify more effectively, such specific benefits of CAVs in mixed-stream traffic, it is essential to understand and simulate the behavior of HDV drivers in such environments. Microscopic traffic flow (MTF) modeling presents an opportunity to carry out this task. Therefore, a fundamental research problem that continues to attract interest is the modelling of microscopic traffic maneuvers including car-following, merging, and lane-changing. Also, MTF models describe the driving behavior, local traffic rules and possible restrictions of the vehicle, which are used to demonstrate collective phenomena such as traffic breakdowns, traffic instabilities, and the propagation of stop-and-go waves. These model models traditionally are calibrated with respect to macroscopic traffic data, for example, flow-occupancy and velocity-occupancy data collected based on loop detectors (Bando et al., 1995). By helping to comprehend the fundamental dynamics of traffic flow, MTF models serve as a powerful approach to assess the impacts of such flow in terms of safety, stability, and efficiency. MTF models can help understand and simulate the behavior of HDV drivers, and thus help assess the impacts of any mitigation efforts in terms of flow safety, stability, and efficiency. As microscopic traffic data have become increasingly available, the feasibility of analyzing and comparing microscopic traffic flow models with real microscopic data has gained traction in the literature (Chen et al., 2010; Huang et al., 2018; Kesting and Treiber, 2008; Punzo et al., 2012; Vieira da Rocha et al., 2015). The performance of car-following models greatly relies on the parameters they contain; and different drivers certainly have various parameters, since these parameters indicate the driver's unique driving habits. Thus, calibration microscopic models with empirical trajectory data can help provide better understanding of human drivers' behavior

In an effort to build on the previous research in this area, the present paper identified a number of research opportunities and therefore seeks to make research contributions in these respects:

- Consideration of mixed (rather than uniform) traffic flow: In the literature, most studies that evaluated the safety and mobility benefits of CAVs were predicated on the assumption of full CAV market penetration (Gunter et al., 2019; Milanes et al., 2014; Schakel et al., 2010). While the findings of these studies represent pioneering efforts and can be considered groundbreaking, the assumption of full market penetration of CAVs, may be rather unrealistic. This is because, in reality, CAVs are likely to be ushered into the market incrementally, and full market penetration is expected to occur only in the distant future after a lengthy transition (Litman, 2019). If this is true, then after their market entry, CAVs will coexist with HDV to form a "mixed" traffic stream, for a long time. Unlike CAVs whose movements can be well planned and designed in advance, HDV movements involve higher unpredictability and heterogeneity, and therefore, uncertainties in mixed traffic environments as explained in an earlier paragraph. A number of studies have argued that in such environments where CAVs share roads with HDVs, the latter will impair the performance of not only CAVs but also the overall traffic stream (Milanés et al., 2013; van den Broek et al., 2011). Thus, to deploy CAVs smoothly in the transition period, new directions of mathematical modeling and control analysis that duly consider HDV-induced uncertainty, are needed.

- Designing of CAV controllers based on empirical data: In previous work, controller frameworks were tested based on some assumed numerical cases (Li et al., 2020). In this paper, however, the HDV dynamics are calibrated with real-world empirical data. The data smoothing process and model calibration are fundamental for implementing the proposed control framework in real time. This provides the CAV controller the flexibility to capture heterogeneity in human driver behavior and to incorporate the effect of such heterogeneity.

- Capture of HDV-CAV interactions: A number of studies have examined CAV impacts on mixed traffic conditions in a simulated environment (Shladover et al., 2001, 2012; VanderWerf et al., 2002). Such past work is intuitive; however, in most of such work, the interactions between CAVs and HDVs were assumed to be the same with those between HDVs. The framework in the present paper accounts for potential HDV-CAV interactions. Since human drivers may have different level of acceptance of CAVs, their car-following behavior may differ from each other. Specifically, drivers who have lower trust in automated systems may tend to adopt conservative levels of headway upon realization that the preceding vehicle is autonomous.

The rest of the paper is organized as follows. Section 2 reviews related work in literature and Section 3 presents the experiment settings and assumptions. Section 4 discusses noise artifacts in NGSIM dataset and the smoothing process prior to using the data for model calibration. Then, details of calibration methodology and results are presented in Section 5. Based on microscopic models calibrated, the controller design problem is formulated as an optimization problem in Section 6, and Section 7 discusses the results and summarizes the study's findings. **Figure 1** presents the flow of the proposed controller design framework.

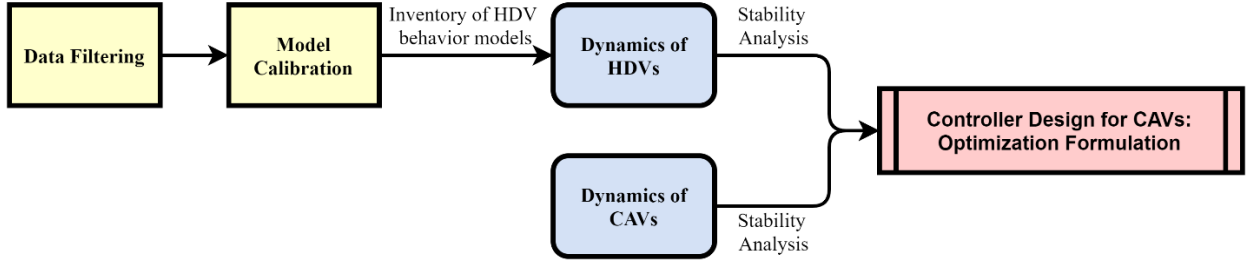

**Figure 1 Controller design framework proposed in this study**
**RELATED WORK**

**Car-following Models under Investigation**
Microscopic traffic models describe the motion of each individual vehicle. In other words, they model actions including individual drivers' accelerations/decelerations as a response to surrounding traffic using an acceleration strategy toward a desired velocity in the free-flow regime, a braking strategy for approaching other vehicles or obstacles, and a car-driving strategy for maintaining a safe distance when driving behind another vehicle. Microscopic traffic models typically assume that human drivers react to the stimulus from neighboring vehicles with the dominant influence originating from the directly leading vehicle known as follow-the-leader or car-following approximation. The dynamics of a standard car-following phenomenon can be expressed as (**Equation 1**):

$$\ddot{x}_i = \frac{d}{dt}x_i = f(\dot{x}_i, \Delta x_i, \Delta \dot{x}_i) \tag{1}$$

where $x_i, \dot{x}_i$ and $\ddot{x}_i$ denote the displacement, velocity and acceleration of the $i$-th vehicle. Based on **Equation 1**, the acceleration of vehicle $i$ depends on its velocity $\dot{x}_i$, headway $\Delta x_i := x_{i-1} - x_i$ and velocity difference, $\Delta \dot{x}_i$, between the preceding vehicle. Intelligent Driver Model (IDM) (Treiber et al., 2006) and Optimal Velocity Model (OVM) (Bando et al., 1995) are widely-used concepts. In this paper, we use OVM to simulate drivers' car-following behaviors.

*Optimal Velocity Model*
The OVM model can be expressed as (**Equation 2**):

$$\ddot{x}_i = \alpha[V(\Delta x_i) - \dot{x}_i] \tag{2}$$

Where: $\alpha$ represents the driver's sensitivity. $V(\Delta x_i)$ denotes an optimal velocity, which is a function of the headways:

$$V(\Delta x_i) = V_0 \big[\tanh m(\Delta x_i - b_f) - \tanh m\,(b_c - b_f)\big] \tag{3}$$

In **Equation 3**, parameters are defined as: $V_0$ represents free flow velocity; $b_c$ denotes the minimum headway and $b_f$ is the headway corresponding to the inflection point; $m$ is a scale value.

The model describes the following behaviors: the driver perceives the gaps and determines an optimal velocity at which the driver desires to travel. However, in most of the cases, there exists a deviation between the optimal and the current velocity. Awareness of such deviation stimulates the driver to reduce the deviation by accelerating or decelerating. There are four parameters to be estimated in this function using empirical data (**Equation 3**), including $V_0, m, b_f$ and $b_c$:

- The maximum velocity for a large enough headway is given by $V_{max} = V_0\big[1 - \tanh m(b_c - b_f)\big]$.

- The headway, $\Delta x = b_f$, corresponds to the inflection point of the Optimal Velocity function, where $V(b_f) = V_{max} - V_0$.
- The Optimal Velocity becomes zero at $\Delta x = b_c$, which is regarded as an effective car length or stopping distance. Thus, it should be a little larger than the average length of vehicles $l_c$: $b_c = l_c + \delta = s_0$ (in the IDM).

However, the original Optimal Velocity (OV) model does not explicitly account for driver response time which could introduce time delay in the HDV's dynamics. Bando et al. (1998)'s modification of the OV model explicitly considers delay, and the system's state-space description can be written as (**Equation 4**):

$$\ddot{x}_i(t) = \alpha_i \left( V(\Delta x_i(t - \tau_i)) - \dot{x}_i(t - \tau_i) \right) \tag{4}$$

As shown in **Equation 4**, due to the natural reaction time of human drivers, the HDV will react to stimuli not at the exact time it receives the stimuli but after several seconds. Additionally, due to heterogeneity in reaction times across the HDV driver population, the sensitivity parameter $\alpha_i$ and the time delay parameter $\tau_i$ are not necessarily the same for different HDVs. Helbing and Tilch (1998) examined this issue. Fitting the model with empirical data, they found that this model may lead to some unrealistic behavior including sharp accelerations and decelerations. To overcome this limitation, (Jiang et al., 2001) proposed a full velocity-difference model (FVDM) (**Equation 5**):

$$\ddot{x}_i(t) = \alpha_i \left( V(\Delta x_i(t - \tau_i)) - \dot{x}_i(t - \tau_i) \right) + \beta_i \Delta v(t - \tau_i) \tag{5}$$

The parameters $\alpha_i$ and $\beta_i$ reflect the relative weights associated with the state of vehicle $i$ in traveling at the optimal velocity and the state of the vehicle in following the preceding vehicle. Hence, calibrating the FVDM model with explicit time delay means the estimation of the following parameters:

- Weights of traveling at optimal velocity $\alpha_i$,
- Weights of following the preceding vehicle $\beta_i$,
- Minimum headways $b_c$ (effective car length),
- Headways corresponding to inflection point $b_f$,
- Free flow velocity $V_0$,
- Distance scale $m$,
- Perception-reaction time delay $\tau_i$.

For IDM, the parameter space for each vehicle $i$ can be expressed as $\boldsymbol{\theta}_i = \{\alpha_i, \beta_i, b_c, b_f, V_0, m, \tau_i\}$

**Autonomous Vehicle Control**

Adaptive cruise control (ACC) is a concept that involves automatic adjustment of a vehicle's cruise-control velocity (in the presence of downstream traffic) to a safe following distance. As a partially automated driving feature, ACC seeks to enable longitudinal control of the vehicle and to reduce the driver's workload (Eskandarian, 2003; Ioannou and Chien, 1993). Over the years, rapid developments in information/communication technologies have yielded promising extensions of the ACC concept to a cooperative system feature known as CACC (cooperative adaptive cruise control). CACC systems leverage the availability of vehicle-to-vehicle (V2V) communications to collect extensive and reliable information (Milanés et al., 2013). Doing this promotes awareness of the surrounding traffic environment and thereby improves the control system's reliability and performance. Previous researchers developed conceptual CACC models to evaluate their feasibility in achieving traffic safety and efficiency (Kato et al., 2002).

**Next Generation SIMulation (NGSIM) Dataset**
To support the development of traffic microsimulation models,  the Federal Highway Administration has supported real-world data collection (**Figure 2**) and simulation algorithm design. The NGSIM program consists of datasets for several corridors –  I-80, Lankershim Boulevard, and US-101 (Alexiadis et al., 2004). The development of these datasets is considered a watermark in the evolution of traffic research as they provide high-quality data validating and calibrating microscopic traffic models and driver behavior models (Thiemann et al., 2008). These datasets contain detailed data on vehicle trajectory for investigating diver behavior.  Synchronized digital video cameras recorded vehicles passing through the study areas and their trajectory data were transcribed from the videos to provide precise locations for each vehicle every one-tenth of a second.

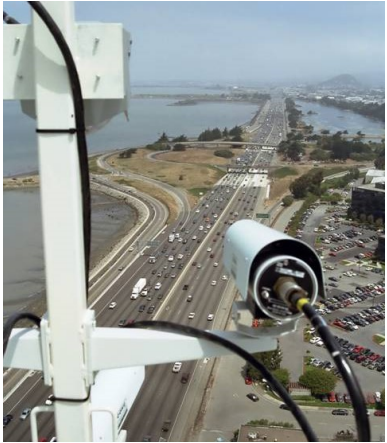

**Figure 2 Recording of vehicle trajectories using a digital video camera overlooking a highway** (Halkias and Colyar, 2006)**.**

A number of studies have used the NGSIM dataset. Roess and Ulero studied trends and sensitives in weaving sections (Roess and Ulerio, 2007); Zhang and Kovvali (2007) and Goswami and Bham (2007) assessed gap acceptance theory, and Toledo and Xohar (2007) investigated lane-change behavior. Very recently, attention has focused on using the data to facilitate CAV implementation to help reduce congestion, increase safety and improve productivity, and increase the capacity of existing road facilities (Talebpour and Mahmassani, 2016). Understanding the microscopic behavior of human drivers has become essential not only because of mixed traffic stream conditions but also because of the need to design intelligent CAVs. For example, Zhu et al.  (2020) and Zhang et al. (2019) designed controllers for AVs by training car-following events using information from the NGSIM dataset. Rather than directly using the longitudinal and lateral position information from the trajectories, other studies have used vehicle velocities and accelerations in microscopic modeling (Chen et al., 2010; Kesting and Treiber, 2008; Thiemann et al., 2008). Specifically, Thiemann, Treiber and Kesting mentioned that the NGSIM dataset (Thiemann et al., 2008) may contain noise artifacts which is further discussed at the subsequent section of the present paper.

**Calibration Methods**
*Data Filtering*
The original trajectory data provided by NGSIM was acquired through digital video processing methods and therefore contains some noise. In cases where velocities and accelerations play a significant role, such as testing or calibrating car-following models or lane-changing models, the noise in the positional information (longitudinal and lateral information) is greatly increased and a direct application is not possible. The NGSIM dataset contains instantaneous velocity and acceleration of each vehicle. As shown in Figure3, the raw data exhibits some noise artifacts. Consider, for example, the data of the US-101 during the 7:50-8:05 AM period with its velocity and acceleration distributions (**Figure 3**). For this snapshot of the dataset, large accelerations are conspicuously frequent (accelerations exceeding $\pm 3$ m/s$^2$ constitute over 11% of all the reported accelerations). This suggests that the drivers were engaged in very frequent and very sharp accelerations and decelerations for no obvious reason, which is not realistic. Additionally, there exist several spikes in the velocity distribution. In **Figure 4**, the example trajectory suggests that the driver is changing between hard acceleration and hard deceleration several times a second, which also, is unrealistic. Thus, it is reasonable infer that these unrealistic dynamics of velocities and accelerations are caused by errors in data processing such as discretization errors. Therefore, a filtering process, which smooths data based on weighted averages of neighboring observed data, is needed prior to the calibration. Exponential smoothing is widely used for smoothing time series data, for example, trajectory data. Thiemann et al. adopted a symmetric exponential moving average filter (sEMA) to all the trajectories (Thiemann et al., 2008) before using the data for calibration.

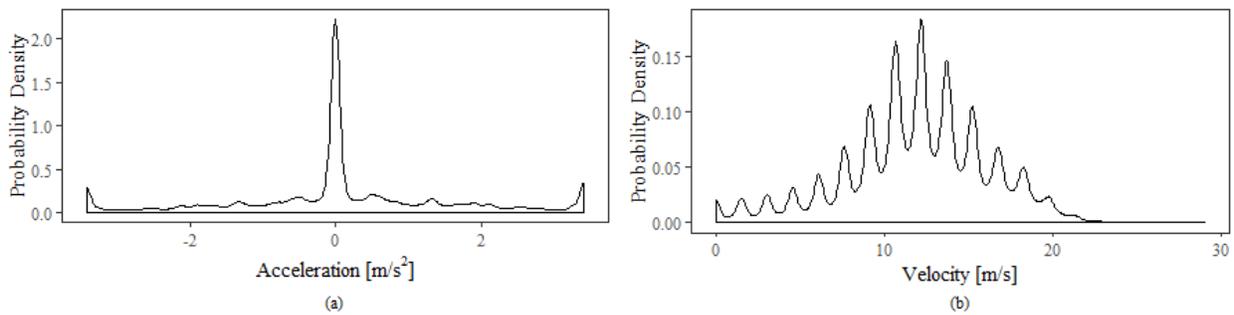

**Figure 3 Distributions of the raw data in US-101 dataset during the time period of 7:50-8:05 a.m.: (a) acceleration distribution, (b) velocity distribution.**

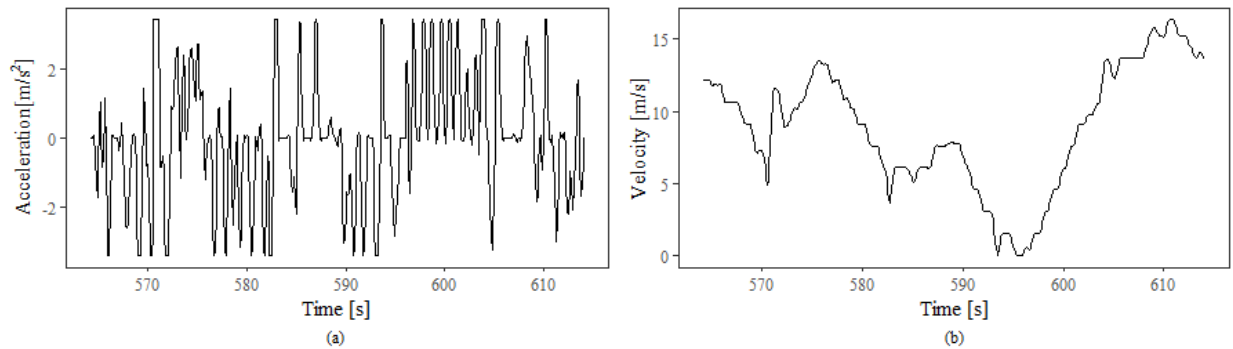

**Figure 4 Example trajectory of vehicle ID = 1989 (a) acceleration dynamics of example trajectory (vehicle ID = 1989), (b) velocity dynamics of example trajectory.**

*Genetic Algorithm (GA)*
Genetic Algorithm (GA) is a heuristic nonlinear optimization algorithm inspired by the process of biological evolution and is capable of generating high-quality solutions to nonlinear problems. This algorithm has been used successfully in several research domains (Holland, 2019) and in the context of this study, there exist a few  studies that used GA to estimate parameters to calibrate microscopic models with empirical trajectory data (Chen et al., 2010; Hoogendoorn and Hoogendoorn, 2010; Kesting and Treiber, 2008). The calibration process seeks to minimize the difference between the measured (data-based) driving behavior and the simulated behavior based on the car-following model under consideration. We seek to minimize headway errors using operators as mutation, crossover and selection.

**EXPERIMENT SETTINGS**

The standard schema of a platoon (**Figure 5**) comprises a leading vehicle (vehicle 0) and mixed traffic (both connected human-driven vehicles (HDVs) and connected and autonomous vehicles (CAVs)) that follow (or, are upstream of) the leading vehicle. The leading vehicle, which is an HDV, exhibits unpredictable movements that is the source of the string instability.

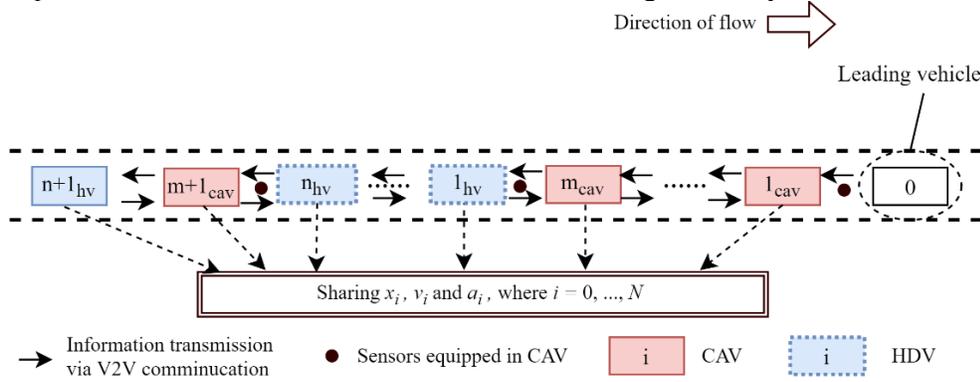

**Figure 5 Standard platoon schema in a mixed traffic stream.**

To ensure a rigorous inquiry into the issue, we make the following assumptions for a platoon: (a) All the vehicles in the traffic stream (HDVs and AVs) are well-connected; (b) CAVs have no communication error or propagation time delay. The first assumption can be ensured when defining platoons that the $N$ vehicles in the platoon should be within communication range. The second assumption can be considered realistic because there exist promising innovative technologies, such as 5G, that facilitate such efficiency in vehicle-to-vehicle communications. Therefore, it is expected that the issue of communication error and delay can be addressed.

**DATA PREPARATION**

As discussed in previous sections, the original values of velocity and acceleration included in the NGSIM dataset suffer from noise artifacts and the data were filtered prior to calibration of the car-following models. All the dataset measurements were collected with a given time discretization interval $dt = 0.1$ s. Therefore, the continuous time variable $t$ can be represented as $t = kdt$. Let $x_i(t)$ denote the measured position of vehicle $i$ at time $t$. In discrete system, $x_i(t)$ can be replaced by $x_i(k)$, where $k$ represents time step and $k \in (1, \ldots, N_i)$ ($N_i$ is the total number of time steps recorded of vehicle $i$). The sEMA filter use an exponential kernel (**Equation 6**):

$$g(t) = \exp\left(-\frac{|t|}{T}\right) \tag{6}$$

$T$ is the smoothing time width which is the only parameter in sEMA filter. In equivalent, **Equation 7** is:

$$g(k) = \exp\left(-\frac{|k|}{\Delta}\right) \tag{7}$$

Where: $\Delta = \frac{T}{dt}$ is the corresponding smoothing time steps.

The smoothed position of vehicle $i$ at time step $k$ is given by **Equation 8**:

$$\bar{x}_i(k) = \frac{1}{Z}\sum_{j=k-D}^{k+D} x_i(j)e^{-\frac{|k-j|}{\Delta}} \tag{8}$$

Where: $Z = \sum_{j=k-D}^{k+D} e^{-\frac{|k-j|}{\Delta}}$ is the normalization constant.

The smoothing window width $D$ for each time step $k$ is determined by **Equation 9**:

$$D = min\{3\Delta, k-1, N_i - k\} \tag{9}$$

Based on **Equation 9**, the smoothing window width is chosen to be three times the smoothing time steps and to ensure symmetry, it would decrease for the points near the trajectory boundaries. Besides the fundamental smoothing mechanism, order of differentiations and parameter choosing, selecting proper smoothing time width $T$ need to be determined. Thiemann, et al found that first the differentiation to velocities and accelerations and then the smoothing of the three variables turned out to better reproduce the original trajectories (Thiemann et al., 2008). Also, the smoothing times for positions, velocities and accelerations were chosen as $T_x = 0.5$ s, $T_v = 1$ s and $T_a = 4$ s. Equivalently, $\Delta_x = 5$, $\Delta_v = 10$, and $\Delta_a = 40$. After conducting smoothing process, noise artifacts discussed in previous sections are alleviated (**Figure 6**). The smoothed data are further used to calibrate car-following models and then serve as fundamental elements for designing CAV controllers.

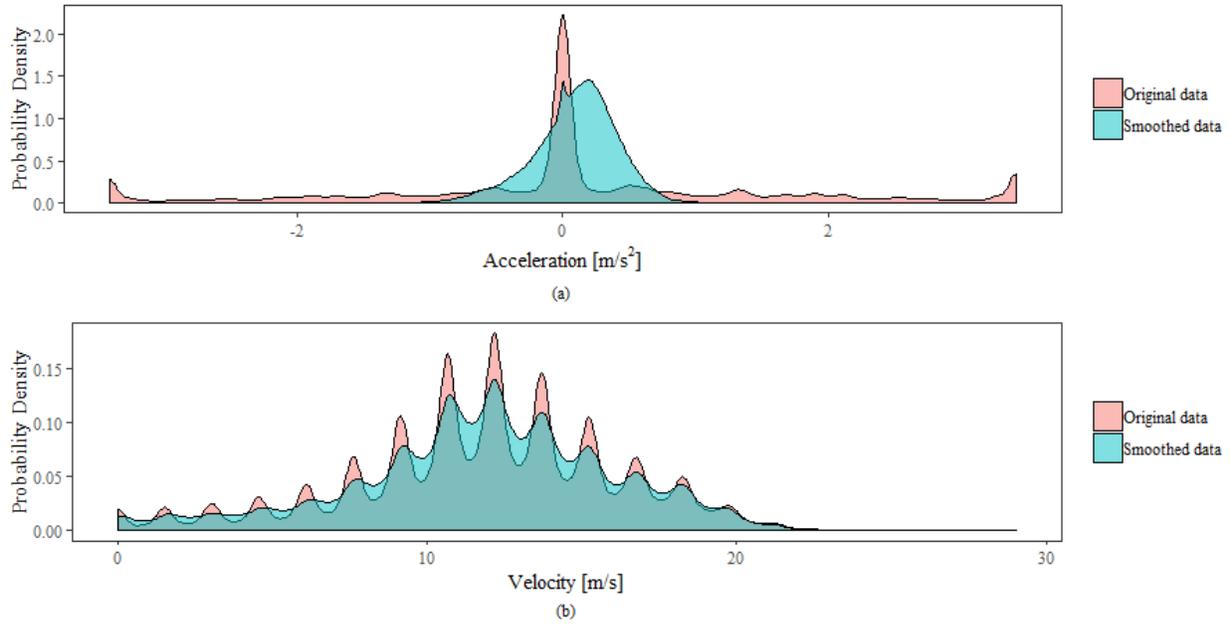

**Figure 6 Comparison of distributions without smoothing: (a) acceleration distribution (b) velocity distribution of the original and smoothed data in the US-101 dataset, 7:50-8:05 AM**

## MODEL CALIBRATION

### Objective Functions
The calibration seeks to minimize the difference between the measured driving behavior based on data and the simulated one based on the car-following model under consideration. Specifically, we attempt to minimize the error in headway which is defined as $s(t) := x_{i-1}(t) - x_i(t)$. Given measured headways, $s^{data}(t) = x_{i-1}^{data}(t) - x_i^{data}(t)$, and simulated headways $s^{sim}(t) = x_{i-1}^{data}(t) - x_i^{sim}(t)$, the difference between these two values is what we minimize (the superscript $data$ denotes the measured value and $sim$ denotes the simulated value). Equivalently, in discrete system where $t = kdt$, $s^{sim}(k) = x_{i-1}^{data}(k) - x_i^{sim}(k)$ As Kesting and Treiber did (Kesting and Treiber, 2008), the error is defined as (Eqn 10):

$$F(s^{sim}) = \sqrt{\frac{1}{\langle|s^{data}|\rangle}\left\langle\frac{(s^{sim}-s^{data})^2}{|s^{data}|}\right\rangle} \tag{10}$$

Where: the operator $\langle s \rangle$ representing the temporal average of a times series of duration $T$ (Eqn 11):

$$\langle e \rangle = \frac{1}{T} \int_0^T e(t) dt \tag{11}$$

Since we are dealing with discrete dataset, **Equation 13** can be rewritten in discretized form as (Eqn 12):

$$\langle e \rangle = \frac{1}{N_a dt} \sum_{k=1}^{N_a} e(k) dt = \frac{1}{N_a} \sum_{k=1}^{N_a} e(k) \tag{12}$$

The mixed errors defined in **Equation 10** combines relative $F_{ref}(s^{sim})$ (**Equation 13**) errors and absolute errors (**Equation 14**).

$$F_{rel}(s^{sim}) = \sqrt{\left\langle \left( \frac{s^{sim} - s^{data}}{s^{data}} \right)^2 \right\rangle} = \sqrt{\frac{1}{N_a} \sum_{k=1}^{N_a} \frac{1}{s^{data}(k)^2} (s^{sim}(k) - s^{data}(k))^2} \tag{13}$$

$$F_{abs}(s^{sim}) = \sqrt{\frac{\left\langle (s^{sim} - s^{data})^2 \right\rangle}{\left\langle s^{data} \right\rangle^2}} \tag{14}$$

The relative errors can be viewed as averaging difference in simulated and measured headways by the inverse of headways, which is sensitive to small headways. While the absolute error is more senstive to large headways, it is inclined to overestimate error when headways are large. The mixed errors combine these two measures to better assess the performance of the model with regard to its data fitting efficacy.

**Optimization Algorithm**

To calibrate the car-following models, data on adjacent vehicle pairs are needed. Thus, after the smoothing process, we paired the preceding and following vehicles that appear in the same time instance. The following vehicles are initialized with $x_i^{sim}(0) = x_i^{data}$, $v_i^{sim}(0) = v_i^{data}(0)$. Given headways, velocity difference and other measurements as inputs, the car-following model parameterized by $\boldsymbol{\theta}$ is used to compute the acceleration. The position and velocity of the following vehicles are updated as follows:

$$x_i^{sim}(k + 1) = x_i^{sim}(k) + v_i^{sim}(k) dt + \frac{1}{2} a_i(k) dt^2 \tag{15}$$

$$v_i^{sim}(k + 1) = v_i^{sim}(k) + a_i(k) dt \tag{16}$$

The genetic algorithm is:

1. Each model with a parameter set is viewed as an "individual", and a "population" consists of $N$ sets of parameters.
2. In each "generation", the fitness of each individual is calculated based on **Equation 10**, mixed error measuremnt.
3. Pairs of two individuals are randomly selected from the current population based on their fitness level. To generate a new individual, the genes of all individuals(their model parameters), are varied randomly corresponding to a mutation that is controlled by a given probability. The resulting new generation is then used in the next iteration.
4. The evolution terminates after convergence, which is specified by a constant best-of-generation score for at least a given number of generations.

To identify a reasonable solution to the nonlinear problem, there should be constraints posed to the parameter space. For FVDM, the weights of traveling at optimal velocity $\alpha_i$ and weights of following the preceding vehicle $\beta_i$ are restricted to the interval [1, 10], the stopping distance $b_c$ to [0.1,8] m, the headways corresponding to the inflection point $b_f$ to [0.1, 100], free flow velocity $V_0$ to [1, 70] m/s, the distance scale $m$ to [0.00001, 10]. In the genetic algorithm, the population size $N = 50$ and the evlution will be terminated when either conditions is satisfied:1) the evolution iterates after 1000 times; 2) the fitness score keeps unchanged for 100 consecutive generations.

**Calibration Results**
For each pair of leading and following vehicle, time frames that both vehicles has trajectory data were chosen. With 0.1 s time step, there are more than 600 data points available for each pair. Additionally, we attempted to take drivers perception-reaction time into consideration and constructed models of HDVs with explict time delay (Li et al., 2020). However, the parameter $\tau_i$ imposes negligible influence on the calibration error. This result is consistent with discoveries by other studies and one possible interpretation is the drivers' ability of anticipating the incoming situation, which considerably compensates for their reaction time (Kesting and Treiber, 2008). We calibrated car-following models for all valid vehicle pairs, which serves as model inventory and will be randomly chosen during controller design process. From the data, the vehicle with ID 1989 is used as an example to show the evolution of GA algorithm and calibration results (**Figure 7**).

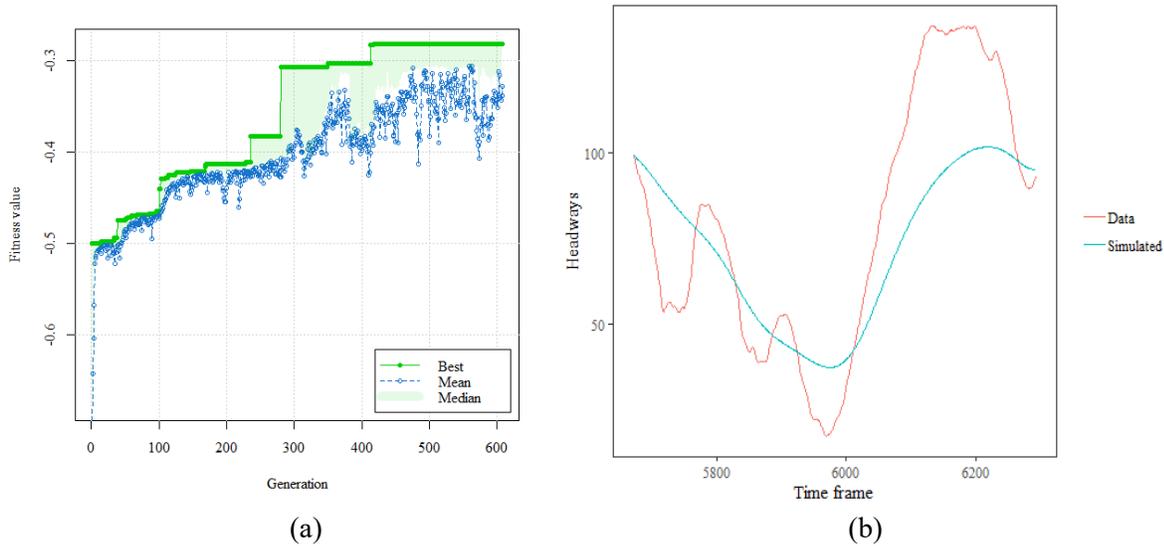

(a)                                            (b)

**Figure 7 (a) Visualization of GA search process to calibrate microscopic car-following models for vehicle 1989 (b) Comparison of actual headways and simulated value.**

# CONTROLLER DESIGN

## Dynamics of the Human-driven Vehicles
The stability of the non-linear system is then analyzed by considering the equilibrium state. When the flow is uniform, all the HDVs travel at the same velocity $v^*$ with desired headways $\Delta x_i^*$. The desired headways represent an encapsulation of the constant time headway rule and the constant clearance rule. A Taylor expansion of the full velocity difference model in the vicinity of the equilibrium point, yields:

$$\ddot{x}_i(t) = \alpha_i V'\big(\Delta x_i^*(t)\big)(\Delta x_i(t) - \Delta x_i^*) - \alpha_i(\dot{x}_i(t) - v^*) + \beta_i \Delta \dot{x}_i(t) \tag{17}$$

The state variables should be bounded. Therefore, the modified variables $\tilde{x}_i(t) := x_i(t) - x_i^*(t)$ were used. Then, **Equation 17** can be rewritten as Equation 21 based on **Equation 18-20**:

$$\Delta x_i^*(t) \coloneqq \lambda_{2i} v_i(t) + \lambda_{3i} \tag{18}$$

$$x_i(0) \coloneqq -\sum \Delta x_i^*(0) = -\sum(\lambda_{2i}v^* + \lambda_{3i}) \quad \forall i \tag{19}$$

$$x_i^*(t) \coloneqq x_i(0) + tv^* \quad \forall i \tag{20}$$

$$\ddot{\tilde{x}}_i(t) = k_{1i}\left(\tilde{x}_{i-1}(t-\tau_i) - \tilde{x}_i(t-\tau_i) - \lambda_{2i}\dot{\tilde{x}}_i(t-\tau_i)\right) - k_{2i}\tilde{x}_i(t-\tau_i) + k_{3i}\left(\dot{\tilde{x}}_{i-1}(t-\tau_i) - \dot{\tilde{x}}_i(t-\tau_i)\right) \tag{21}$$

where:
$k_{1i} = \alpha_i V'(\Delta x_i^*)$, $k_{2i} = \alpha_i$, and $k_{3i} = \beta_i$

After conducting a Laplace transformation of $\tilde{x}_i(t)$ and denoting the states as $\tilde{X}_i(s)$, the transfer function $T_i(s)$ can be derived as follows (**Equations 22 and 23**):

$$s^2\tilde{X}_i(s) = k_{1i}\left(e^{-s\tau_i}\tilde{X}_{i-1}(s) - e^{-s\tau_i}\tilde{X}_i(s) - \lambda_{2i}se^{-s\tau_i}\tilde{X}_i(s)\right) - k_{2i}se^{-s\tau_i}\tilde{X}_i(s) + k_{3i}se^{s\tau_i}\left(\tilde{X}_{i-1}(s) - \tilde{X}_i(s)\right) \tag{22}$$

$$T_i(s) = \frac{\tilde{X}_i(s)}{\tilde{X}_{i-1}(s)} = \frac{(k_{1i}+sk_{3i})e^{-s\tau_i}}{s^2+s(k_{2i}+k_{3i}+k_{1i}\lambda_{2i})e^{-s\tau_i}+k_{1i}e^{-s\tau_i}} \tag{23}$$

**Dynamics of the Connected and Autonomous Vehicles**

For purposes of modeling the CAV dynamics, we use the full velocity-difference model without-time-delay (Jiang et al., 2001) (**Equation 24**):

$$\ddot{x}_i(t) = k_1\left(\Delta x_i(t) - \Delta x_i^*(t)\right) + k_2(\dot{x}_i(t) - v^*) + k_3\Delta\dot{x}_i \tag{24}$$

Similar to the mathematical manipulation made to the HDV model, the variable $\tilde{x}_i(t)$ can be transformed as follows: $\tilde{x}_i(t) = x_i(t) - x_i^*(t)$, and then **Equation 24** can be re-arranged to yield (**Equation 25**):

$$\ddot{\tilde{x}}_i(t) = k_1\left(\tilde{x}_{i-1}(t) - \tilde{x}_i(t) - \lambda_2\dot{\tilde{x}}_i(t)\right) - k_2\dot{\tilde{x}}_i(t) + k_3\left(\dot{\tilde{x}}_{i-1}(t) - \dot{\tilde{x}}_i(t)\right) \tag{25}$$

Therefore, the transfer function for CAVs can be expressed as (**Equation 26**):
$$T_A(s) = \frac{\tilde{X}_i(s)}{\tilde{X}_{i-1}(s)} = \frac{(k_1+sk_3)}{s^2+s(k_2+k_3+k_1\lambda_2)+k_1} \tag{26}$$

**Controller Performance Metrics**

*Vehicular String stability ($\mathcal{L}_2$ stability)*

To analyze the string stability of the mixed traffic platoon, the frequency-domain approach is used. For each HDV $i$, substitute s with $j\omega$, where $\omega$ (frequency), is $\geq 0$. The magnitude-squared frequency response is obtained:

$$|T_i(j\omega)|^2 = \frac{k_{1i}^2+\omega^2 k_{3i}^2}{\omega^2 K^2+\omega^4+k_{1i}^2+f(\omega)} \tag{27}$$

where $K = k_{2i} + k_{3i} + k_{1i}\lambda_{2i}$, $f(\omega) = -2\omega^3 K\sin\omega\tau_i - 2\omega^2 k_{1i}\cos\omega\tau_i$. To ensure the uniform boundedness for the HDVs, $|T(j\omega)| \leq 1$, $\forall \omega \geq 0$. When $\tau_i = 0$ and $\lambda_2 = 0$, the stability requirement yields **Equation 28**:

$$\omega^2 \geq 2k_{1i} - k_{2i}^2 - 2k_{2i}k_{3i} \tag{28}$$

This inequality is consistent with the string stability condition for the linearized dynamics of a non-delayed system (Ioannou and Xu, 1994; Orosz et al., 2010). With these two inequalities being satisfied, the critical frequency for $i$-th HDV model is given by (**Equation 29**):

$$\omega_{i0}^H = \sqrt{2k_{1i} - k_{2i}^2 - 2k_{2i}k_{3i}} \tag{29}$$

Based on the string stability analysis above, the $i$-th HDV is string unstable, given a perturbation with frequency $\omega$, $0 < \omega < \omega_{i0}^H$. For the HDV platoon that consists of multiple HDVs, the most critical frequency is given by (**Equation 30**):

$$\omega_0^H = \min_i \omega_{i0}^H \tag{30}$$

Carrying out a similar variable substitution $s = j\omega$ to CAV model, it is determined that the magnitude-squared frequency response is given by (**Equation 31**):

$$|T_A(j\omega)|^2 = \frac{k_1^2 + \omega^2 k_3^2}{(k_1 - \omega^2)^2 + \omega^2(k_2 + k_3 + k_1\lambda_2)^2} \tag{31}$$

To ensure string stability of CAVs for $\forall \omega$, $k_1, k_2$ and $k_3$ should be tuned to satisfy $|T_A(j\omega)| \leq 1$. This means that (**Equation 32**):

$$\omega^4 + \omega^2(k_2^2 + k_1^2\lambda_2^2 + 2k_2k_3 + 2k_1k_2\lambda_2 + 2k_1k_3\lambda_2 - 2k_1) \geq 0 \tag{32}$$

Simplifying (28) yields the requirement for CAV string stability for all perturbations, which can be expressed as (**Equation 33**):

$$k_2^2 + k_1^2\lambda_2^2 + 2k_2k_3 + 2k_1k_2\lambda_2 + 2k_1k_3\lambda_2 - 2k_1 \geq 0 \tag{33}$$

Therefore, appropriate levels of $k_1$, $k_2$, and $k_3$ should be selected based on **Equation 33**, to ensure the string stability of CAVs. This condition is consistent with the well-known conditions (Wilson and Ward, 2011).

*Platoon string stability ($\mathcal{L}_2$ weak string stability)*
We demonstrate requirements for vehicular stability, but as mentioned in previous section of this paper, the stability of an individual vehicle does not necessarily translate into the stability of the entire stream of vehicles. Thus, we further require platoon string stability, which can be expressed mathematically as **Equation 34**:

$$\left\| \prod_{\forall i} T_i(j\omega) \right\|_\infty \leq 1 \tag{34}$$

Platoon string stability is essential in mixed traffic streams. Specifically, with some unstable oscillations triggered by the uncontrolled leading vehicle (vehicle 0 in **Figure 3**), the AV can dampen the shockwave and thereby mitigate the propagation of unstable waves. Assume that for any perturbation of frequency $\omega$, $\omega \in (0, \omega_0^H)$, the CAV can stabilize the $n$ HDVs that follow it. From **Equation 34**, **Equations 35-38** can be derived:

$$\left| T_A(j\omega) \prod_i^n T_i(j\omega) \right| \leq 1 \tag{35}$$

$$\log|T_A(j\omega)| + \sum_i^n \log|T_i(j\omega)| \leq 0 \tag{36}$$

$$\log|T_A(j\omega)| + \sum_i^{n+1} \log|T_i(j\omega)| > 0 \tag{37}$$

The maximum number of stabilized HDVs, $n^*_{stable}$, under $\forall \omega \in (0, \omega_0^H)$, is given by:

$$n^*_{stable} = \min_{\omega} \min_{n} \left\{ \left(log|T_A(j\omega)| + \sum_i^n log|T_i(j\omega)|\right) \cdot \left(log|T_A(j\omega)| + \sum_i^{n+1} log|T_i(j\omega)|\right) \leq 0 \right\} \qquad (38)$$

*Safety Considerations*
The achievement of string stability does not guarantee that the system is collision-free. Additionally, it does not eliminate extremely conservative and inefficient conditions (which occur when the CAV attempts to maintain a large headway). Therefore, besides using a frequency-domain approach, it is important to impose constraints on the headway (Wu et al., 2018). Then, there exist two headway-related safety considerations:

- Minimum headway $\Delta x^-$: This ensures that the vehicle is free from collision and is equal to an effective vehicle length (Bando et al., 1998). Consider the traffic condition where each driver maintains an extra distance margin that might be needed for stopping to avoid collision. Then, the effective vehicle length exceeds the actual vehicle length.
- Maximum headway $\Delta x^+$: This ensures that the vehicle will not maintain an unreasonably large headway because that will result in low throughput and therefore, traffic inefficiency.

Based on the above considerations, the headway constraints can be represented as:

$$\Delta x^- \leq \Delta x_i \leq \Delta x^+ \qquad (39)$$

Then, for a specific disturbance with magnitude $\beta$, Wu (2018) showed that given the headway constraints, the number of HDVs that a single AV can stabilize is (**Equation 40**):

$$n^*_{s/e} = \min_{\omega} \{ \min_{n} \{ (log|1 - T_A(j\omega)| + \sum_i^n log|T_i(j\omega)| - log(\eta)) \cdot log|1 - T_A(j\omega)| + \sum_i^{n+1} log|T_i(j\omega)| - log(\eta) \leq 0 \} \} \qquad (40)$$

Where: $\eta = \Delta/\beta$, represents the relative scale of the disturbance.

**Optimization Problem Formulation**
The overall controller design problem for CAVs can be formulated as a multi-objective problem with the following objectives: (a) maximize the number of HDVs that the CAV can stabilize given oscillations which may trigger stop-and-go waves, (b) minimize the risks of collision. The underlying settings of the problem can be summarized as:

- The problem considers that there exist certain conditions under which the HDVs are not string stable. This means that small perturbations from a uniform flow are amplified as they propagate from the leading vehicle to vehicles that follow it.
- Human-driven vehicle models are calibrated based on NGSIM dataset. The parameters that characterize human behaviors are not necessarily the same across individual human drivers.
- All the HDVs that are to be stabilized by the CAV are well-connected through electronic connectivity. In this paper, this assumption can be considered appropriate because in the study, we define this based on a specific range of vehicle-to-vehicle communication

The optimization problem can then be expressed as (**Equations 41-44**):

$$max \; n^*_{stable}, n^*_{s/e} \qquad (41)$$

s.t.,

$$T_A(s) = \frac{(k_1 + sk_3)}{s^2 + s(k_2 + k_3 + k_1\lambda_2) + k_1} \tag{42}$$

$$k_2^2 + k_1^2\lambda_2^2 + 2k_2k_3 + 2k_1k_2\lambda_2 + 2k_1k_3\lambda_2 - 2k_1 \geq 0 \tag{43}$$

$$k_1, k_2, k_3 \geq 0 \tag{44}$$

where $k_1, k_2, k_3$ are decision variables to be tuned.

By tuning the parameters that characterize the controller, $k_1$, $k_2$ and $k_3$, the maximum number of HDVs that can be safely stabilized, is optimized at $k_1 = 0$, $k_2 \approx \eta k_3$. **Figure 8** presents the relationship between the number of HDVs that can be stabilized by CAVs and controller gains, and based on it, the optimal controller parameter can be determined.

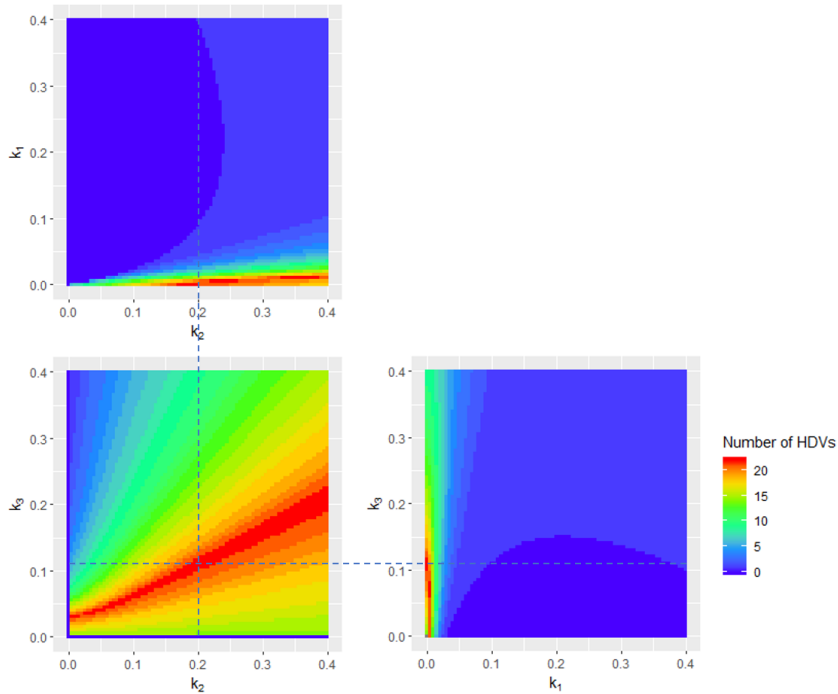

**Figure 8 Heatmaps of relationships between controller gains parameter combinations and the maximum number of HDVs that can be stabilized**

## DISCUSSIONS AND CONCLUSIONS

In this paper, we design controllers for CAVs in mixed traffic flow conditions to stabilize a traffic platoon. With the prospective duality of operations of CAVs and human driven vehicles in that they will share the same road space, it is essential for CAVs to capture the dynamics of HDVs and based on the knowledge of surrounding traffic. For CAVs to understand the behavior of surrounding HDVs, microscopic car-following models for the HDVs were calibrated using their trajectory data. Since the original trajectory data provided by NGSIM dataset contains some noises and the information related to vehicle velocities and accelerations cannot be directly used in model calibration, and therefore the data were filtered prior to calibration. The HDV dynamics are to be calibrated in real time. Also, human drivers may have different level of acceptance of CAVs, their car-following behavior may differ from each other. Specifically, drivers who have lower trust in autonomous mobility may tend to adopt a conservative headway when they find the preceding vehicle is autonomous vehicle. Therefore, the control design framework proposed in this paper presents flexibility to capture heterogeneity in human driver behavior and to incorporate the effect of such flexibility, and to model potential interactions between HDVs and CAVs. Further, the proposed framework for the CAV controller design is shown to be capable of calibrating noisy trajectory data and determining optimal control parameter for CAV to mitigate stop-and-go waves triggered by the HDV. Serving as an extension of previous research the framework is not restricted to numerical experiment in this paper but can be extended to other application areas. Such flexibility is possible partly because the controller was designed based on real-world trajectory data.

This study has a few limitations that could serve as a beacon not only for practitioners wishing to adopt the study results for implementation but also for future researchers wishing to replicate the study and subsequently to improve the model further. In the paper, we assumed there is no communication time delay for the CAV models. However, in reality, there could exist a little delay in the information processing by CACC controllers. In addition, the calibration errors can be further minimized by either testing with more advanced car-following models. Finally, future research could compare models calibrated from other trajectory datasets and explore the transferability of the microscopic models.

## ACKNOWLEDGMENTS

This work was supported by Purdue University's Center for Connected and Automated Transportation (CCAT), a part of the larger CCAT consortium, a USDOT Region 5 University Transportation Center funded by the U.S. Department of Transportation, Award #69A3551747105. The contents of this paper reflect the views of the authors, who are responsible for the facts and the accuracy of the data presented herein, and do not necessarily reflect the official views or policies of the sponsoring organization.

## AUTHOR CONTRIBUTIONS

The authors confirm contribution to the paper as follows: all authors contributed to all sections. All authors reviewed the results and approved the final version of the manuscript.